\documentclass[aps,tightenlines,showpacs]{revtex4}%

\usepackage{amsfonts}
\usepackage{amsmath}
\usepackage{amssymb}
\usepackage{graphicx}

\begin{document}

\title[Two-way and three-way negativities]{Two-way and three-way
negativities of three qubit entangled states}
\author{S. Shelly Sharma}
\email{shelly@uel.br}
\affiliation{Depto. de F\'{\i}sica, Universidade Estadual de Londrina, Londrina
86051-990, PR Brazil }
\author{N. K. Sharma}
\email{nsharma@uel.br}
\affiliation{Depto. de Matem\'{a}tica, Universidade Estadual de Londrina, Londrina
86051-990 PR, Brazil }
\thanks{}
\keywords{Global negativity, K-way negativity, three qubit entanglement}

\begin{abstract}
In this letter we propose to quantify three qubit entanglement using global
negativity along with $K-$way negativities, where $K=2$ and $3$. The
principle underlying the definition of $K-$way negativity for pure and mixed
states of $N-$subsystems is PPT sufficient condition. However, $K-$way
partial transpose with respect to a subsystem is defined so as to shift the
focus to $K-$way coherences instead of $K$ subsystems of the composite
system.
\end{abstract}

\maketitle

Quantum entanglement is not only a fascinating aspect of multipartite
quantum systems, but also a physical resource needed for quantum
communication, quantum computation and information processing in general.
Bipartite entanglement is well understood, however, many aspects of
multipartite entanglement are still to be investigated. Peres \cite{pere96}
and the Horedecki \cite{horo96,horo197,horo297} have shown a positive
partial transpose (PPT) of a bipartite density operator to be a sufficient
criterion for classifying bipartite entanglement. Negativity \cite%
{zycz98,vida00} based on Peres Horodecski criterion has been shown to be an
entanglement monotone \cite{eise01,vida02,plen05}. Negativity is a useful
concept being related to the eigenvalues of partially transposed state
operator and can be calculated easily. In this letter, we define $2-$way and 
$3-$way negativities and propose a classification of three qubit states
based on measures related to global, $2-$way and $3-$way negativities.
General definition of $K-$way negativities for pure and mixed states of $N-$%
subsystems is given in ref \cite{shel06}. The $K-$way partial transpose with
respect to a subsystem is defined so as to shift the focus to $K-$way
coherences instead of $K$ subsystems of the composite system. While pure $K-$
partite entanglement of a composite system is generated by $K-$way
coherences, $K-$partite entanglement can, in general, be present due to $%
\left( K-1\right) $ way coherences as well.

\section{Global Negativity}

The Hilbert space, $C^{d}=C^{d_{1}}\otimes C^{d_{2}}\otimes C^{d_{3}},$
associated with a quantum system composed of three qubits is spanned by
basis vectors $\left\vert i_{1}i_{2}i_{3}\right\rangle ,$ where $i_{m}=0$ or 
$1,$ and $m=1,2,3$. Here $d_{m}=2$ is the dimension of Hilbert space
associated with $m^{th}$ qubit. To simplify the notation we denote the
vector $\left\vert i_{1}i_{2}i_{3}\right\rangle $ by $\left\vert
\prod\limits_{m=1}^{3}i_{m}\right\rangle $and write a general three qubit
pure state as 
\begin{equation}
\widehat{\rho }=\sum_{\substack{ i_{1},i_{2},i_{3},  \\ j_{1},j_{2},j_{3}}}%
\left\langle \prod\limits_{m=1}^{3}i_{m}\right\vert \widehat{\rho }%
\left\vert \prod\limits_{m=1}^{3}j_{m}\right\rangle \left\vert
\prod\limits_{m=1}^{3}i_{m}\right\rangle \left\langle
\prod\limits_{m=1}^{3}j_{m}\right\vert .  \label{1}
\end{equation}%
To measure the overall entanglement of a subsystem $p,$ we shall use twice
the negativity as defined by Vidal and Werner \cite{vida00}, and call it
global Negativity $N_{G}^{p}$. It is an entanglement measure based on
Peres-Hororedecki \cite{pere96,horo96} NPT (Negative partial transpose)
sufficient criterion for classifying bipartite entanglement. The global
partial transpose of $\widehat{\rho }$ (Eq.(\ref{1})) with respect to
sub-system $p$ is defined as%
\begin{equation}
\widehat{\rho }_{G}^{T_{p}}=\sum_{\substack{ {i}_{1},i_{2},i_{3},  \\ %
j_{1},j_{2},j_{3}}}\left\langle j_{p}\prod\limits_{m=1,m\neq
p}^{3}i_{m}\right\vert \widehat{\rho }\left\vert
i_{p}\prod\limits_{m=1,m\neq p}^{3}j_{m}\right\rangle \left\vert
\prod\limits_{m=1}^{3}i_{m}\right\rangle \left\langle
\prod\limits_{m=1}^{3}j_{m}\right\vert .  \label{3}
\end{equation}%
The partial transpose $\widehat{\rho }_{G}^{T_{p}}$ of an entangled state is
not positive. Global Negativity, defined as%
\begin{equation}
N_{G}^{p}=\left\Vert \rho _{G}^{T_{p}}\right\Vert _{1}-1,  \label{4}
\end{equation}%
is an entanglement monotone and measures the entanglement of subsystem $p$
with its complement in a bipartite split of the composite system. Global
negativity vanishes on ppt-states and is equal to the entropy of
entanglement on maximally entangled states.

\section{$K-$way partial transpose}

The $K-$way partial transpose of\ $\rho $ with respect to subsystem $p$ is
obtained by transposing the indices of subsystem $p$ in those matrix
elements, $\left\langle \prod\limits_{m=1}^{3}i_{m}\right\vert \widehat{\rho 
}\left\vert \prod\limits_{m=1}^{3}j_{m}\right\rangle $, that satisfy the
condition $\sum\limits_{m=1}^{3}\left\vert j_{m}-i_{m}\right\vert =K$, where 
$K=0$ to $3$.

A typical matrix element of the three qubit state operator $\widehat{\rho }$
involves a change of state of $K$ subsystems, where $K=0$ to $3$ . For
example, a matrix element involving a change of state of two subsystems
looks like $\left\langle i_{1}i_{2}i_{3}\right\vert \widehat{\rho }%
\left\vert j_{1}j_{2}i_{3}\right\rangle $ ($i_{1}\neq j_{1},i_{2}\neq j_{2}$%
). The set of two distinguishable subsystems that change state while one of
the sub-systems does not, can be chosen in three distinct ways. In general,
the number of spins that are flipped to get a vector $\left\vert
j_{1}j_{2}j_{3}\right\rangle $ from the vector $\left\vert
i_{1}i_{2}i_{3}\right\rangle $ is $K=\sum\limits_{m=1}^{3}\left\vert
j_{m}-i_{m}\right\vert $. The operator $\widehat{\rho }$ can be split up
into parts labelled by $K$ ($0\leq K\leq 3$) and written as 
\begin{equation}
\widehat{\rho }=\sum_{K=0}^{3}\widehat{R}_{K},  \label{5}
\end{equation}%
with%
\begin{equation}
\widehat{R}_{K}=\sum_{I_{K}}\left\langle
\prod\limits_{m=1}^{3}i_{m}\right\vert \widehat{\rho }\left\vert
\prod\limits_{m=1}^{3}j_{m}\right\rangle \left\vert
\prod\limits_{m=1}^{3}i_{m}\right\rangle \left\langle
\prod\limits_{m=1}^{3}j_{m}\right\vert .  \label{6}
\end{equation}%
Here $I_{K}=\left\{ i_{1},i_{2},i_{3},j_{1},j_{2},j_{3}:\left(
\sum_{m}\left\vert j_{m}-i_{m}\right\vert \right) =K\right\} $. The $2-$way
and $3-$way partial transpose with respect to qubit $p$ are defined as%
\begin{eqnarray}
\widehat{\rho }_{2}^{T_{p}} &=&\sum_{K=0,1,3}\widehat{R}_{K}+  \notag \\
&&\sum_{\text{ }I_{2}}\left\langle j_{p}\prod\limits_{\substack{ m=1  \\ %
m\neq p}}^{3}i_{m}\right\vert \widehat{\rho }\left\vert i_{p}\prod\limits 
_{\substack{ m=1  \\ m\neq p}}^{3}j_{m}\right\rangle \left\vert
\prod\limits_{m=1}^{3}i_{m}\right\rangle \left\langle
\prod\limits_{m=1}^{3}j_{m}\right\vert  \label{7}
\end{eqnarray}%
and%
\begin{eqnarray}
\widehat{\rho }_{3}^{T_{p}} &=&\sum_{K=0}^{2}\widehat{R}_{K}+  \notag \\
&&\sum_{I_{3}}\left\langle j_{p}\prod\limits_{\substack{ m=1  \\ m\neq p}}%
^{3}i_{m}\right\vert \widehat{\rho }\left\vert i_{p}\prod\limits_{\substack{ %
m=1  \\ m\neq p}}^{3}j_{m}\right\rangle \left\vert
\prod\limits_{m=1}^{3}i_{m}\right\rangle \left\langle
\prod\limits_{m=1}^{3}j_{m}\right\vert .  \label{9}
\end{eqnarray}

\begin{figure}[t]
\centering \includegraphics[width=3.75in,height=5.0in,angle=-90]{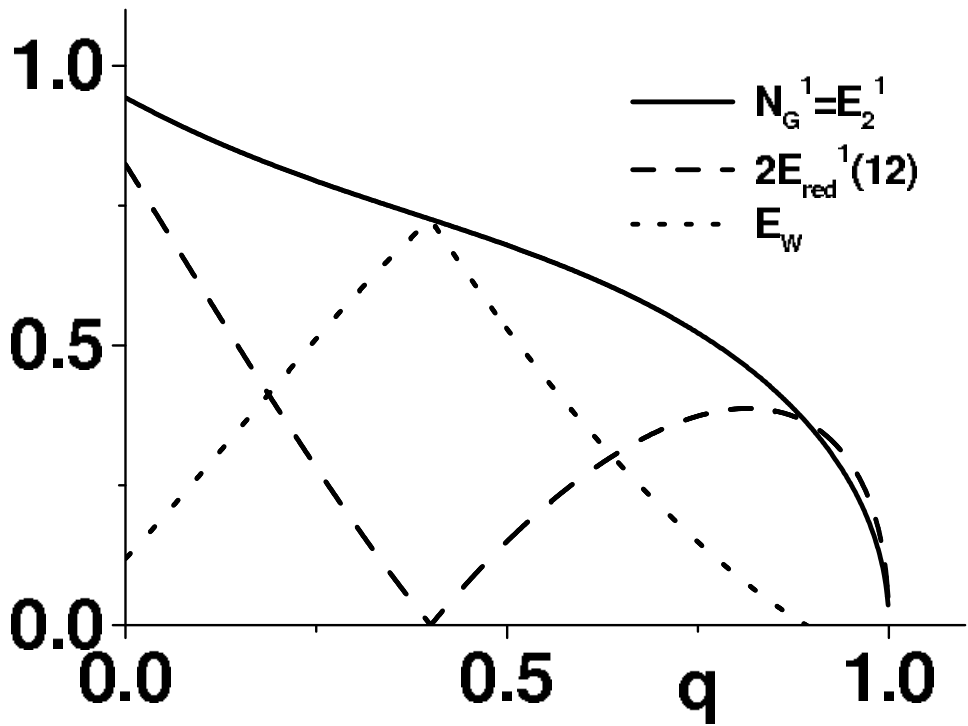}
\caption{ $N_{G}^{1}$, $2E_{red}^{1}(12)$ and $E_{W}$ versus parameter $q$
for the state $\hat{\protect\rho}_{1}$.}
\label{fig1}
\end{figure}

\section{K-way Negativity}

The $K-$way negativity calculated from $K-$way partial transpose of matrix $%
\rho $ with respect to subsystem $p$, is defined as 
\begin{equation}
N_{K}^{p}=\left\Vert \rho _{K}^{T_{p}}\right\Vert _{1}-1,  \label{10}
\end{equation}%
where $\left\Vert \rho _{K}^{T_{p}}\right\Vert _{1}$ is the trace norm of $%
\rho _{K}^{T_{p}}$. Using the definition of trace norm and the fact that $%
tr(\rho _{K}^{T_{p}})=1$, we get$\left\Vert \rho _{K}^{T_{p}}\right\Vert
_{1}=2\sum_{i}\left\vert \lambda _{i}^{K-}\right\vert +1$, $\lambda
_{i}^{K-} $ being the negative eigenvalues of matrix $\rho _{K}^{T_{p}}$.
The negativity, $N_{K}^{p}=2\sum_{i}\left\vert \lambda _{i}^{K-}\right\vert $
($p=1,2,3$), depends on $K-$ way coherences and is a measure of all possible
types of entanglement attributed to $K-$ way coherences. Intuitively, for a
system to have pure $N-$partite entanglement, it is necessary that $N-$way
coherences are non-zero. On the other hand, $N-$partite entanglement can be
generated by $(N-1)-$ way coherences, as well. For a three qubit system,
maximally entangled tripartite GHZ state is an example of pure tripartite
entanglement involving $3-$way coherences. The global negativity $%
N_{G}^{p}=N_{3}^{p}=1$, for maximally entangled three qubit GHZ state.
Maximally entangled W-state is a manifestation of tripartite entanglement
due to $2-$way coherences. For pure states as well as those mixed states for
which the density matrix is positive, entanglement of a subsystem is
completely determined by global Negativity $N_{G}^{p}$ and the hierarchy of
negativities $N_{K}^{p}$ ($K=2,...N),$ calculated from $\rho _{K}^{T_{p}}$
associated with the $p^{th}$ sub-system. For three qubit system, $N_{2}^{p}$%
, $N_{3}^{p}$, and $N_{G}^{p}$ ( $p=1,2,3$) quantify the coherences present
in the composite system.

\section{How much bi and tripartite entanglement is generated by 2-way and
3-way negativities?}

A natural question is, how much of global negativity comes from $2-$way
transpose and how much has its origin in $3-$way transpose, for a given
qubit? The operator $\widehat{\rho }_{G}^{T}$ in its eigen basis is written
as 
\begin{equation}
\widehat{\rho }_{G}^{T_{p}}=\sum\limits_{i}\lambda _{i}^{G+}\left\vert \Psi
_{i}^{G+}\right\rangle \left\langle \Psi _{i}^{G+}\right\vert
+\sum\limits_{i}\lambda _{i}^{G-}\left\vert \Psi _{i}^{G-}\right\rangle
\left\langle \Psi _{i}^{G-}\right\vert ,  \label{11}
\end{equation}%
where $\lambda _{i}^{G+}$and $\left\vert \Psi ^{G+}\right\rangle $ ($\lambda
_{i}^{G-}$and $\left\vert \Psi ^{G-}\right\rangle $) are the positive
(negative) eigenvalues and eigenvectors, respectively. As such the
negativity of $\widehat{\rho }_{G}^{T_{p}}$ is given by 
\begin{equation}
N_{G}^{p}=-2\sum\limits_{i}\left\langle \Psi _{i}^{G-}\right\vert \widehat{%
\rho }_{G}^{T_{p}}\left\vert \Psi _{i}^{G-}\right\rangle
=2\sum\limits_{i}\left\vert \lambda _{i}^{G-}\right\vert \text{.}  \label{12}
\end{equation}%
It is easily verified that 
\begin{equation}
\widehat{\rho }_{G}^{T_{p}}=\widehat{\rho }_{2}^{T_{p}}+\widehat{\rho }%
_{3}^{T_{p}}-\widehat{\rho }.  \label{13}
\end{equation}%
Substituting Eq. (\ref{13}) in Eq. (\ref{12}) and recalling that $\widehat{%
\rho }$ is a positive operator with trace one, we get%
\begin{eqnarray}
N_{G}^{p} &=&-2\sum\limits_{i}\left\langle \Psi _{i}^{G-}\right\vert 
\widehat{\rho }_{2}^{T_{p}}\left\vert \Psi _{i}^{G-}\right\rangle
-2\sum\limits_{i}\left\langle \Psi _{i}^{G-}\right\vert \widehat{\rho }%
_{3}^{T_{p}}\left\vert \Psi _{i}^{G-}\right\rangle ,  \notag \\
&=&E_{2}^{p}+E_{3}^{p}\text{,}  \label{14}
\end{eqnarray}%
where $E_{K}^{p}=-2\sum\limits_{i}\left\langle \Psi _{i}^{G-}\right\vert 
\widehat{\rho }_{K}^{T_{p}}\left\vert \Psi _{i}^{G-}\right\rangle $ is the
contribution of $K-$way partial transpose to $N_{G}^{p}$. 
\begin{figure}[t]
\centering \includegraphics[width=3.75in,height=5.0in,angle=-90]{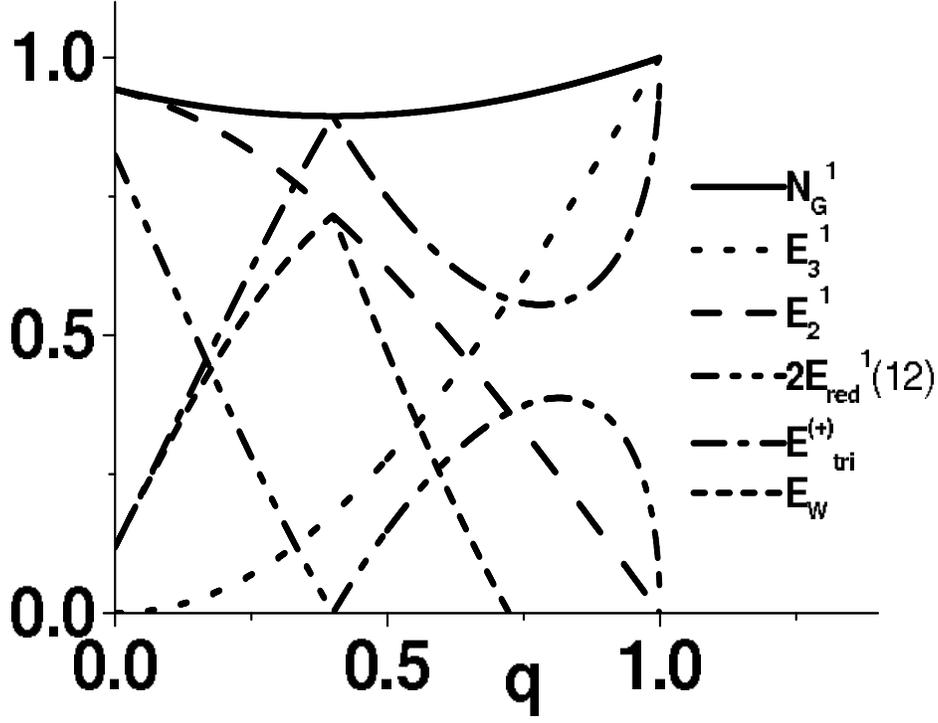}
\caption{ $N_{G}^{1}$, $E_{3}^{1}$, $E_{2}^{1}$, $2E_{red}^1(12)$ $E_{tri}^{(+)}$
and $E_{W}$ as a function of parameter $q$ for the state $\Psi _{2}^{(+)}$.}
\label{fig2}
\end{figure}

The set of states that can be transformed into each other by local unitary
operations lie on the same orbit and have the same entanglement as the
canonical state expressed in terms of the minimum number of independent
vectors \cite{cart99,cart00}. Construction of pure three qubit canonical
state has been given by Acin et al \cite{acin00,acin01}. It is easily
verified that for the states reducible to the canonical state by local
unitary operations, although $N_{G}^{p}$ is invariant under local
operations, $\sum\limits_{p=1}^{3}N_{K}^{p}$ \ varies under local unitary
operations. For a canonical state, $N_{3}=\sum_{p=1}^{3}${\Huge \ }$%
N_{3}^{p} $, is found to lie at a minimum with respect to local unitary
rotations applied to any of the three qubits. We conjecture that for a
canonical state,\ $\min (E_{3}^{1},E_{3}^{2},E_{3}^{3})$ is a measure of
genuine $3-$way entanglement of three qubit system.

Bipartite entanglement of qubit one with qubit two equals the negativity $%
E_{red}^{1}(12)$ of $\widehat{\rho }_{red}^{T_{1}}(12)$ where the reduced
operator, $\widehat{\rho }_{red}(12)=tr_{3}(\widehat{\rho }).$ In case no
W-like tripartite entanglement is present, the bipartite entanglement of a
qubit is given by $E_{red}^{1}(12)+E_{red}^{1}(13)$. For a canonical state,
the measure $E_{2}^{p}$ contains information about the W-like as well as
pairwise entanglement of qubit $p.$

\begin{figure}[t]
\centering \includegraphics[width=3.75in,height=5.0in,angle=-90]{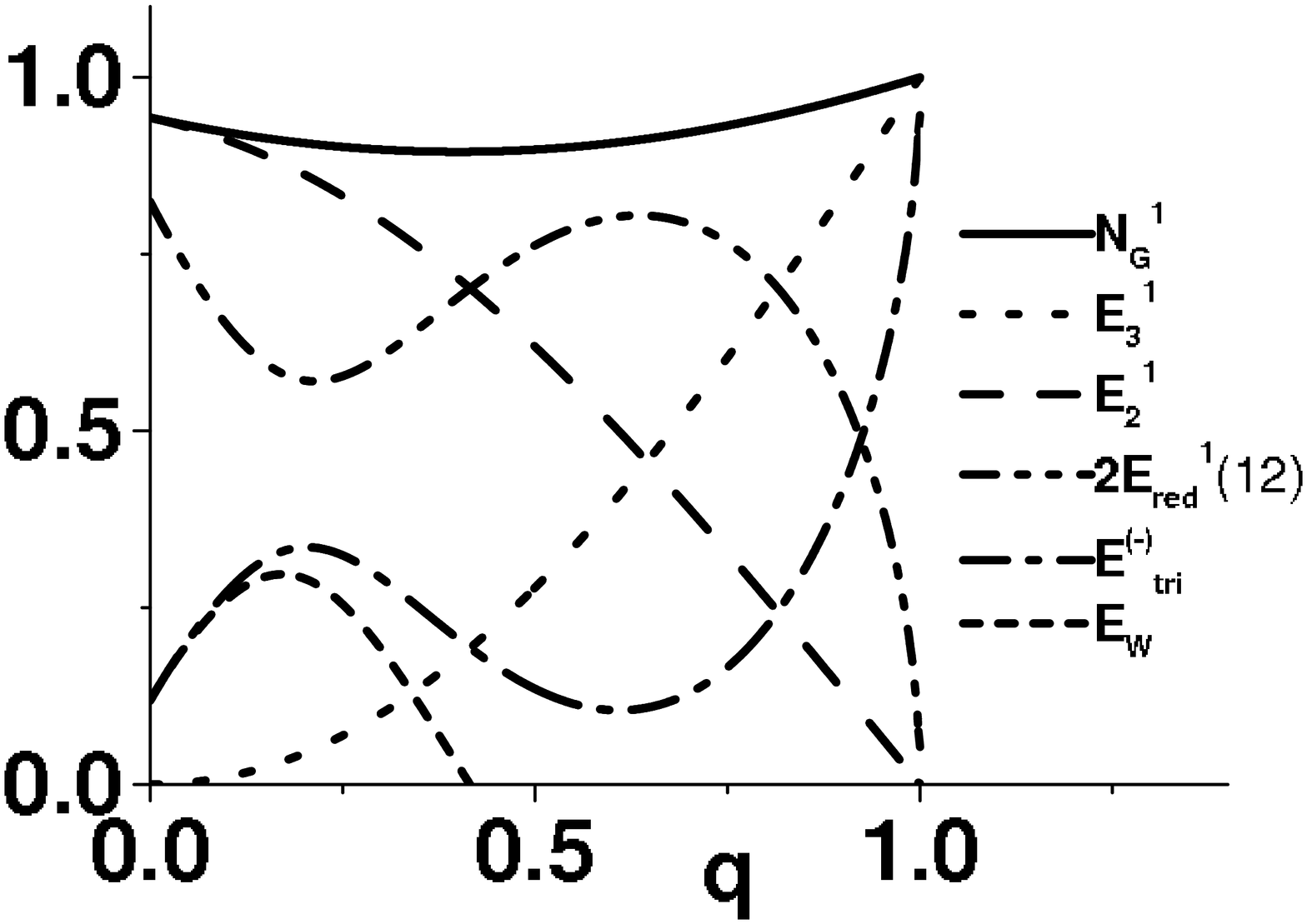}
\caption{ $N_{G}^{1}$, $E _{3}^{1}$, $E_{2}^{1}$, $2E_{2}^1(12)$ $E_{tri}^{(-)}$
and $E_{W}$ as a function of parameter $q$ for the state $\Psi_{2}^{(-)}$.}
\label{fig3}
\end{figure}

\section{Three qubit GHZ and W-type states}

Three qubit Greenberger-Horne-Zeilinger state 
\begin{equation}
\Psi _{GHZ}=\frac{1}{\sqrt{2}}\left( \left\vert 000\right\rangle +\left\vert
111\right\rangle \right) ,\quad \widehat{\rho }_{GHZ}=\left\vert \Psi
_{GHZ}\right\rangle \left\langle \Psi _{GHZ}\right\vert  \label{19}
\end{equation}%
is a maximally entangled state having genuine tripartite entanglement. For
this state $E_{2}^{p}=0$, and $E_{3}^{p}=N_{G}^{p}=1.0$, for $p=1,2,3$. On
the other hand there exists a class of tripartite states akin to maximally
entangled W-state given by 
\begin{equation}
\ \Psi _{W}=\frac{1}{\sqrt{3}}\left( \left\vert 100\right\rangle +\left\vert
010\right\rangle +\left\vert 001\right\rangle \right) ,\quad \widehat{\rho }%
_{W}=\left\vert \Psi _{W}\right\rangle \left\langle \Psi _{W}\right\vert .
\label{20}
\end{equation}%
For the pure state $\Psi _{W}$, $E_{2}^{p}=N_{G}^{p}=0.94$, for $p=1,2,3$.
Bipartite entanglement of qubit one and two in the state $\Psi _{W}$ is
measured by the negativity of partially transposed reduced density operator $%
\widehat{\rho }_{red}(12)$ ($\widehat{\rho }_{red}(12)=tr_{3}(\widehat{\rho }%
_{W}))$, which is found to be $E_{red}^{1}(12)=0.41$ $.$ The total pairwise
entanglement of a qubit in W-state is twice the value of $E_{red}^{p}(12)$ (=%
$0.82)$ and is less than $E_{2}^{p}$($=0.94$). The residue accounts for the
W-type tripartite entanglement of the system and generates $%
E_{W}=N_{G}^{p}-2E_{red}^{p}(12)=0.12$.

\begin{figure}[t]
\centering \includegraphics[width=3.75in,height=5.0in,angle=-90]{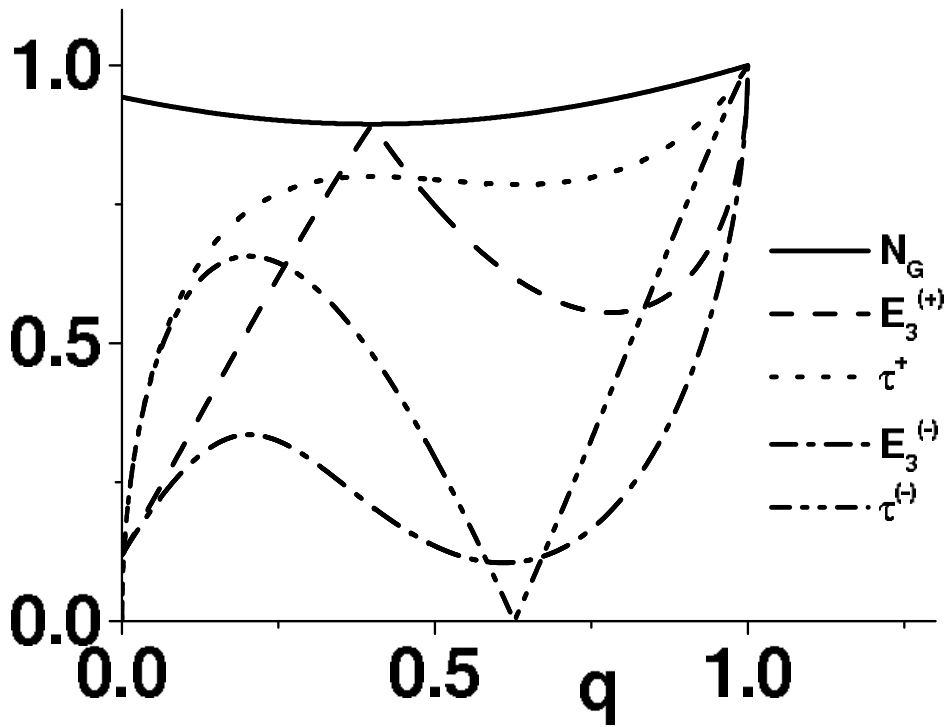}
\caption{ $N_{G}^{1}$, $E_{3}^{+}$, $\protect\tau^{+}$, $E_{3}^{-}$, and $%
\protect\tau^{-}$ as a function of parameter $q$ for the states $%
\Psi_{2}^{(+)}$ and $\Psi_{2}^{(-)}$.}
\label{fig4}
\end{figure}

The three qubit state 
\begin{eqnarray}
\widehat{\rho }_{1} &=&\frac{q}{2}\left\vert 111\right\rangle \left\langle
111\right\vert  \notag \\
&&+\left\vert \sqrt{\frac{q}{2}}\left\vert 000\right\rangle +\sqrt{(1-q)}%
\Psi _{W}\right\rangle \left\langle \sqrt{\frac{q}{2}}\left\vert
000\right\rangle +\sqrt{(1-q)}\Psi _{W}\right\vert  \label{21}
\end{eqnarray}%
has no genuine tripartite entanglement and no $3-$way coherences as such $%
N_{G}^{1}=N_{2}^{1}$. Analogous to the case of state $\Psi _{W}$, we have $%
E_{red}(12)=2E_{red}^{1}(12)$ and $E_{W}=N_{G}^{1}-2E_{red}^{1}(12)$. Fig. 1
displays $N_{G}^{1},$ $2E_{red}^{1}(12)$ and $E_{W}$ as a function of
parameter $q$.

To decipher the interplay of genuine tripartite entanglement and
entanglement generated by $2-$way coherences, we examine the coherences of
single parameter pure states 
\begin{equation}
\Psi _{2}^{(+)}=\sqrt{q}\Psi _{GHZ}+\sqrt{(1-q)}\Psi _{W},\quad \widehat{%
\rho }_{2}^{+}=\left\vert \Psi _{2}^{(+)}\right\rangle \left\langle \Psi
_{2}^{(+)}\right\vert \qquad 0\leq q\leq 1  \label{22}
\end{equation}%
and 
\begin{equation}
\Psi _{2}^{(-)}=\sqrt{q}\Psi _{GHZ}-\sqrt{(1-q)}\Psi _{W},\quad \widehat{%
\rho }_{2}^{-}=\left\vert \Psi _{2}^{\left( -\right) }\right\rangle
\left\langle \Psi _{2}^{(-)}\right\vert \qquad 0\leq q\leq 1.  \label{23}
\end{equation}%
For a given value of $q$, the state may have bipartite,\ genuine tripartite
as well as W-type entanglement as seen by $N_{G}^{1},E_{3}^{1},$ and $%
E_{2}^{1}$ displayed in Figs. (2) and (3). Bipartite entanglement ($%
E_{red}(12)=2E_{red}^{1}(12)$ ) of a qubit in a state $\widehat{\rho }%
_{2}=\left\vert \Psi _{2}\right\rangle \left\langle \Psi _{2}\right\vert $
is obtained by calculating the negativity $E_{red}^{1}(12)$ of partially
transposed $\widehat{\rho }_{red}(12)$ defined as $\widehat{\rho }%
_{red}(12)=Tr_{3}(\widehat{\rho }_{2})$. We may remark that the states $%
\widehat{\rho }_{1}$ and $\widehat{\rho }_{2}^{+}$ share the same $\widehat{%
\rho }_{red}(12)$ with $E_{red}^{1}(12)$ going to zero at $q=0.4$. Figs. (2)
and (3) also display total bipartite entanglement of a qubit, $%
2E_{red}^{1}(12)$, total tripartite entanglement $E_{tri}^{\left( \pm
\right) }=\left( N_{G}^{1}\right) ^{\pm }-2E_{red}^{1\pm }(12)$ and W-state
like entanglement $E_{W}^{1}=N_{G}^{1}-E_{3}^{1}-2E_{red}^{1}(12)$, of a
qubit in the states $\Psi _{2}^{(+)}$ and $\Psi _{2}^{(-)}$, respectively.
Tripartite entanglement $E_{tri}^{\left( \pm \right) }$, resulting from the
interference of $N_{W}^{1}$ and $N_{3}^{1}$ shows that $2-$way coherence is
partially annihilated by the $3-$way coherence and vice versa. Wooters three
tangle calculated from concurrence \cite{woot98,coff00} has been used to
study tripartite entanglement of three qubit system in ref. \cite{lohm06}.
Three tangle for states $\Psi _{2}^{(\pm )}$, calculated from

\begin{equation*}
\tau ^{(\pm )}=\left\vert q^{2}\pm \frac{8\sqrt{6(q(1-q)^{3}}}{9}\right\vert
,
\end{equation*}%
and tripartite entanglement $E_{tri}^{\left( \pm \right) }$ are plotted as a
function of $q$ in Fig. 4. The two measures show similar trend and match at $%
q=1$, but in general $3-$tangle overestimates the tripartite entanglement of
the three qubit composite system in comparison with $E_{tri}^{\left( \pm
\right) }$.

In summary, we have extended the use of PPT criteria to the case of
tripartite entanglement by splitting the partially transposed three qubit
state operator into matrices that contain information either about bipartite
or about genuine tripartite entanglement. The set of $2-$way, $3-$way and
global negativities make clear distinction between the coherences and nature
of entanglement present in three qubit composite states. The three qubit
canonical states may be classified using $%
N_{G}^{p},E_{2}^{p},E_{3}^{p},E_{red}^{p}(pq)$ and $E_{W}^{p}$ ($p,q=1$ to $3
$, $p\neq q$) as labels. These entanglement measures are easy to calculate
using standard diagonalization routines. We believe that the relations
between $K-$way negativities and entanglement of a composite quantum system,
a) with sub-system dimensions higher than two, and  b) with more than three
subsystems, can be figured out. We expect our results to help advance the
investigation of multipartite entanglement.

\section{Acknowledgements}

Financial support from National Council for Scientific and Technological
Development (CNPq), Brazil and State University of Londrina, (Faep-UEL),
Brazil is acknowledged.


\begin{thebibliography}{99}
\bibitem{pere96} A. Peres, Phys. Rev. Lett. 77, 1413 (1996).

\bibitem{horo96} M. Horodecki, P. Horodecki, and R. Horodecki, Phys. Lett. A
223, 8 (1996).

\bibitem{horo197} P. Horodecki, Phys. Lett. A 232, 333 (1997).

\bibitem{horo297} M. Horodecki, P. Horodecki, and R. Horodecki, Phys. Rev.
Lett. 78, 574 (1997).

\bibitem{zycz98} K. Zyczkowski, P. Horodecki, A. Sanpera, and M.
Lewen-stein, Phys. Rev. A 58, 883 (1998).

\bibitem{vida00} G. Vidal, J. Mod. Opt. 47, 355 (2000).

\bibitem{eise01} J. Eisert, PhD thesis (University of Potsdam, February
2001).

\bibitem{vida02} G. Vidal and R. F. Werner, Phys. Rev. Vol. 65, 032314
(2002).

\bibitem{plen05} M.B. Plenio, Phys. Rev. Lett. 95, 090503 (2005).

\bibitem{shel06} S. Shelly Sharma and N. K. Sharma, preprint
quant-ph/0608062.

\bibitem{cart99} H. A. Carteret, N. Linden , S. Popescu, and A. Sudbery,
Foundations of Physics, Vol. 29, No. 4, 527 (1999).

\bibitem{cart00} H. A. Carteret, A. Higuchi, A. Sudbery, J. Math. Phys. 41,
7932 (2000).

\bibitem{acin00} A. Acin, A. Andrianov, L. Costa, E. Jane, J.I. Latorre, R.
Tarrach, Phys. Rev. Lett. 85, 1560 (2000).

\bibitem{acin01} A Ac\'{\i}n, A Andrianov, E Jan\'{e} and R Tarrach, J.
Phys. A: Math. Gen. 34 6725-6739 (2001).

\bibitem{woot98} W. K. Wootters, Phys. Rev. Lett. 80, 2245 (1998).

\bibitem{coff00} V. Coffman, J. Kundu, and W. K. Wootters, Phys. Rev. A 61,
052306 (2000).

\bibitem{lohm06} R. Lohmayer, A. Osterloh, J. Siewert, and A. Uhlmann,
preprint quant-ph/0606071 (2006).
\end{thebibliography}
\end{document}